\documentclass[a4paper,11pt]{article}
\usepackage{jheppub}
\usepackage[english]{babel}

\usepackage{xcolor}
\usepackage{mathtools}
\usepackage{amsthm}
\usepackage{comment}
\usepackage{braket}
\usepackage{pstool}
\usepackage{soul}
\usepackage{float}
\usepackage{esint}
\usepackage{braket}

\hypersetup{
  colorlinks = true,
  linkcolor  = blue!70!black,
  urlcolor   = red!70!black,
  citecolor  = green!55!black,
  anchorcolor= blue!70!black
}

\newcommand{\figref}[1]{Fig.~\ref{#1}}

\DeclareMathAlphabet\mathbfcal{OMS}{cmsy}{b}{n}
\DeclareSymbolFont{bbold}{U}{bbold}{m}{n}
\DeclareSymbolFontAlphabet{\mathbbold}{bbold}

\DeclareMathOperator{\var}{Var}
\DeclareMathOperator{\im}{Im}
\DeclareMathOperator{\re}{Re}

\title{Euclidean volume fluctuations in de Sitter quantum gravity}

\author[a]{David Blanco,}
\author[a]{Guillem P\'erez-Nadal}
\author[a,b]{and Bruno Sivilotti}

\affiliation[a]{Universidad de Buenos Aires, Departamento de F\'isica and IFIBA - CONICET\\
1428 Buenos Aires, Argentina}
\affiliation[b]{
Perimeter Institute for Theoretical Physics\\
31 Caroline Street North, Waterloo, ON N2L2Y5, Canada.}
\emailAdd{dblanco@df.uba.ar, guillem@df.uba.ar, bsivilotti@perimeterinstitute.ca}

\abstract{The Euclidean formulation of quantum gravity can be interpreted in terms of a probability distribution over Riemannian manifolds. In the context of de Sitter gravity, the statistics of the total volume according to this distribution is encoded in the dependence of the partition function on the cosmological constant. We use this observation to obtain a probability distribution for the volume from known results and proposals for the de Sitter partition function, in several levels of approximation: saddle point, one loop, an all-loop and a non-perturbative proposal in 3 dimensions, and an exact result in 2 dimensions, in the context of Liouville theory. In all cases we find a reasonable behavior: in the classical limit the distribution concentrates around the classical volume, and it spreads as quantum effects are turned on. We also find as a common trend that, as quantum effects are increased, the probability distribution favors increasingly smaller universes.}

\begin{document}

\maketitle

\newpage

\section{Introduction}

The quantization of de Sitter (dS) gravity (namely gravity with a positive cosmological constant and dS asymptotics) is regarded as especially puzzling for several reasons, such as the difficulty to embed it into string theory, the lack of a satisfactory holographic description, and some interpretational issues related to the horizon entropy, see for example \cite{Witten:2001kn,Anninos:2012qw,Galante:2023uyf} and references therein.
The purpose of this paper is to learn something new about dS quantum gravity by extracting some new information from the corresponding partition function.

According to Gibbons and Hawking \cite{gibbons_1977_action}, the gravitational partition function is given by the Euclidean path integral
\begin{equation}\label{Z}
    Z=\int{\mathcal D}g\,e^{-I[g]},
\end{equation}
where one formally sums over all Riemannian manifolds with prescribed asymptotics, and $I$ is the Euclidean action, generically including a contribution from the partition function $Z_m$ of the matter fields,
\begin{equation}\label{action}
    I[g]=-\frac{1}{16\pi G}\int d^dx\sqrt{g}(R-2\Lambda)-\log Z_m[g].
\end{equation}
In the case of dS gravity, the prescribed asymptotics is that of Euclidean dS, namely the sphere. Since the sphere has no asymptotic region, the path integral (\ref{Z}) in this case is a sum over compact Riemannian manifolds without boundary.

A peculiarity of the dS partition function is that it is completely fixed by the coupling constants of the theory (namely $G$, $\Lambda$ and the matter couplings). The inverse temperature $\beta$ does not enter as a free parameter, unlike what happens in AdS and asymptotically flat gravity. One way to understand this is as follows. Asymptotically dS spacetimes have no energy, because their spatial sections have no boundary and energy is a boundary term in General Relativity. Thus, the dS partition function should possibly be regarded as a microcanonical 
partition function with the energy fixed to zero, hence the absence of a free parameter. According to this interpretation, $Z$ is the number of microstates compatible with the dS asymptotics or, in other words, the exponential of the entropy. The saddle-point approximation to (\ref{Z}) then gives the Bekenstein-Hawking formula for the entropy, as we will review below. We should note, however, that this state-counting interpretation (or, in fact, any thermodynamic interpretation) is challenged by one-loop corrections, because these corrections generically introduce imaginary factors in $Z$ \cite{Polchinski:1988ua}. These factors ultimately come from the fact that the Euclidean action (\ref{action}) is not bounded from below, which makes the path integral (\ref{Z}) ill-defined. A resolution to this puzzle, which involves introducing an observer in the description, has been recently proposed in \cite{maldacena_2025_real,Chen:2025jqm}.

In the canonical ensemble, the freedom to vary $\beta$ can be used to extract other information from the partition function besides the entropy, namely the statistics of the energy (expectation value, variance and all other moments). Given that we do not have this freedom in dS gravity (and of course the statistics of the energy is trivial, because all its moments vanish), what other information can be extracted from the dS partition function besides the entropy?

To answer this question, let us think of $e^{-I}/Z$ as a formal probability distribution on Riemannian manifolds. The analogous distribution in ordinary QFT can be used to compute the correlation functions of the theory via analytic continuation. In the gravity case the interpretation of this distribution is less clear, but still it seems worthwhile to study its properties. Our observation is that the $\Lambda$-dependence of $Z$ encodes the statistics of the total volume ${\mathcal V}[g]=\int d^dx\sqrt{g}$ according to this distribution. Indeed, the expectation value and the variance are respectively a first and a second derivative,
\begin{align}\label{variance}
\begin{aligned}
&\langle{\mathcal V}\rangle=\frac{1}{Z}\int{\mathcal D}g\, e^{-I[g]}\,{\mathcal V}[g]=-8\pi G\,\partial_\Lambda\log Z\\
&\var({\mathcal V})=\langle{\mathcal V}^2\rangle-\langle{\mathcal V}\rangle^2=(8\pi G)^2\,\partial_{\Lambda}^2\log Z,
\end{aligned}
\end{align}
and higher derivatives give the rest of the moments. In fact, the probability distribution for the volume is also related in a simple way to the partition function,
\begin{equation}\label{prob}
    p(V)=\frac{1}{Z}\int{\mathcal D}g\, e^{-I[g]}\,\delta\left({\mathcal V}[g]-V\right)=\frac{1}{2\pi}\int_{-\infty}^\infty dt\, e^{-itV}\frac{Z(\Lambda-it8\pi G)}{Z(\Lambda)},
\end{equation}
as one can easily show by writing the delta function as a Fourier integral.
The purpose of this paper is to study this probability distribution, and to see what the known results on $Z$ tell us about it.

Note that the total volume is diffeomorphism-invariant, a necessary condition for a quantity to be an observable in quantum gravity. 
Unfortunately, however, we do not have a satisfactory interpretation of this variable from a physical, i.e., Lorentzian point of view. In the case of the sphere, giving the volume is the same as giving the radius, which in turn gives the area of the dS horizon. Thus, we may speculate that the fluctuations of the Euclidean volume correspond to fluctuations of the horizon in dS quantum gravity. Another comment we want to make is that studying the statistics of the total Euclidean volume would not make sense in asymptotically flat and AdS gravity, because in these cases, due to the asymptotic conditions, the Riemannian manifolds involved in the Euclidean path integral all have infinite volume{\footnote{We note, however, that in the context of Einstein manifolds there is a notion of renormalized volume, see for example \cite{anderson2001bf2,yang2008renormalized}}}.

The paper is organized as follows. In section \ref{sec:basics} we review the basic facts about the geometry and thermodynamics of de Sitter space. In section \ref{sec:conditions} we discuss the conditions that the partition function must satisfy in order for $p$ (given by (\ref{prob})) to be a true probability distribution (i.e., to be positive and normalized) and to have support in the positive reals, as it should because the volume is positive definite. In section \ref{sec:examples} we compute the probability distribution using known results for $Z$ in several different levels of approximation: saddle point, one loop, an all-loop and a non-perturbative proposal in 3 dimensions, and an exact result in 2 dimensions, in the context of Liouville theory. Finally, we close with a discussion in section \ref{discussion}.

\section{De Sitter basics}\label{sec:basics}

De Sitter space is the maximally symmetric solution to vacuum Einstein's equations with $\Lambda >0$. In the so-called global coordinates, the metric is
 \begin{align}\label{global}
    ds^2 = \ell^2 (-d\tau^2+\cosh^2 \tau\hspace{3pt}d\Omega^2_{d-1})\qquad\ell^2 = \frac{(d-2)(d-1)}{2\Lambda}\,.
\end{align}
From this metric it is clear that a constant $\tau$ slice is a closed manifold, a $(d-1)$-sphere with radius $\ell\cosh \tau$. So, for $\tau >0$ this metric describes an exponentially expanding universe.

\begin{figure}
    \centering
    \includegraphics[width=0.4\linewidth]{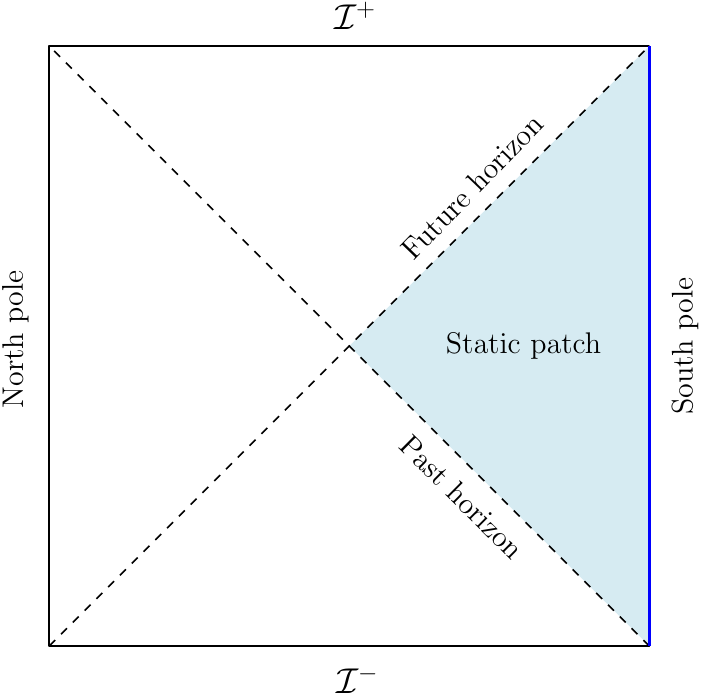}
    \caption{Penrose diagram of de Sitter space. Shaded is the region for which an inertial observer (worldline in blue) has complete causal access.}
    \label{fig:Penrose}
\end{figure}

Because of this exponential expansion, an inertial observer cannot access the full spacetime. The region for which they have full causal access, the ``static patch'', is bounded by a cosmological horizon, as shown in \figref{fig:Penrose}. 
The static patch can be covered by the so-called static coordinates, in which the metric takes the form
\begin{align}\label{static}
    \begin{aligned}
     ds^2 =  -\left(1-\frac{r^2}{\ell^2}\right)dt^2+\frac{dr^2}{1-\frac{r^2}{\ell^2}}+r^2d\Omega^2_{d-2}.
    \end{aligned}
    \end{align}
The observer resides in $r=0$ (blue line in \figref{fig:Penrose}), and the cosmological horizon is at $r=\ell$. 

Gibbons and Hawking observed that this cosmological horizon, like the black hole horizon, has an entropy and a temperature \cite{gibbons_1977_cosmological, gibbons_1977_action}. Let us briefly review these facts using the Euclidean formalism. 
If we Wick-rotate the static metric \eqref{static} by setting $t= -it_E$, we find that, to avoid a conical singularity at $r=\ell$,
the Euclidean time has to be identified as $t_E \sim t_E+2\pi \ell$.  This implies that the cosmological horizon has temperature
\begin{align}
    T = \frac{1}{2\pi \ell}.
\end{align}
Note that, with this identification, the Wick-rotated static metric is that of the sphere of radius $\ell$. Wick-rotating (\ref{global}) gives the same result.

To obtain the entropy, let us compute the gravitational path integral (\ref{Z}) in pure gravity and in the saddle-point approximation,
\begin{equation}
    Z\approx e^{-I_{cl}},
\end{equation}
where by the classical action $I_{cl}$ we mean the smallest on-shell action.  
Any solution $g$ of the vacuum Einstein equations satisfies $R = \frac{2d}{d-2}\Lambda$, so its action is proportional to the volume,
\begin{align}\label{on shell action}
    I[g] = -\frac{1}{16\pi G}\int d^dx \sqrt{g}\hspace{1pt}\frac{4\Lambda}{d-2} = -\frac{\Lambda {\mathcal V}[g]}{4\pi G(d-2)}.
\end{align}
Since $\Lambda>0$, the smallest action corresponds to the largest volume. As it turns out \cite{witten_2025_introduction}, the solution of largest volume is the sphere (i.e., Euclidean dS), so the classical volume is
\begin{equation}
    V_{cl}=\Omega_{d}\ell^d\qquad \Omega_d= \frac{2 \pi^{\frac{d+1}{2}}}{\Gamma\left(\frac{d+1}{2}\right)}
\end{equation}
and hence
\begin{align}\label{action(area)}
    I_{cl} =  -\frac{\Lambda \Omega_{d}\ell^d}{4\pi G(d-2)} = -\frac{(d-1) \Omega_{d}\ell^{d-2}}{8\pi G} =-\frac{\Omega_{d-2}\ell^{d-2}}{4G} = -\frac{A}{4G},
\end{align}
where $A$ is the area of the cosmological horizon. Therefore, the saddle-point approximation to the Euclidean gravitational path integral gives
\begin{align}\label{Z_saddle}
    Z \approx e^{A/4G}.
\end{align}
Interpreting this gravitational path integral as the partition function of the gravitating system, and recalling that dS has zero energy (because its spatial sections have no boundary),
we can directly relate the path integral with the entropy as $S = \log Z$. So we finally get that the dS entropy is given by the Bekenstein-Hawking formula,
\begin{align}
    S = \frac{A}{4G}.
\end{align}
The saddle-point approximation is expected to be accurate in the limit $G\to 0$, although there is no rigorous justification of this because the gravitational action is not bounded from below.

\section{Consistency conditions}\label{sec:conditions}

As explained in the Introduction, from the dS partition function $Z$ one can extract a probability distribution $p$ for the Euclidean volume, Eq.~(\ref{prob}). In terms of the Fourier transform
\begin{equation}
    \phi(t)=\int_{-\infty}^\infty dV e^{itV}p(V),
\end{equation}
the relation between $Z$ and $p$ can be rewritten as
\begin{equation}\label{characteristic}
    \phi(t)=\frac{Z(\Lambda-it8\pi G)}{Z(\Lambda)}.
\end{equation}
A probability distribution is of course positive and normalized, and, since the volume is positive, this distribution in particular should have support in the positive reals. What do these conditions tell us about the partition function? To answer this question, we take $p$ to be a generic tempered distribution, see what each of these conditions says about its Fourier transform $\phi$, and then translate it into a condition on $Z$ via (\ref{characteristic}).

\subsubsection*{Positivity}

If $p$ is positive, then $\phi$ is what is called a positive-definite function, meaning that, for each collection $t_1,t_2,\dots$ of real numbers, the matrix $M_{ij}=\phi(t_i-t_j)$ is positive definite. Indeed, for any vector $v$ we have
\begin{equation}
    \sum_{i,j}M_{ij}v_{i}^*v_j=\int_{-\infty}^\infty dV\,p(V)\,|\tilde v(V)|^2\ge 0,
\end{equation}
where $\tilde v(V)=\sum_j e^{-it_jV}v_j$. In fact, the converse is also true: if $\phi$ is positive definite, then $p$ is positive. This is Bochner's theorem (theorem IX.9 in \cite{reed1975ii}). Thus, the positivity condition is equivalent to saying that the function $\phi$ defined in (\ref{characteristic}) is positive definite. We do not explore this result any further because we will use another one, which assumes support in the positive reals and is easier to handle, see below.

\subsubsection*{Normalization}

Obviously, $p$ is normalized if and only if $\phi(0)=1$. This condition is automatically satisfied by (\ref{characteristic}), so it does not provide non-trivial information about the partition function.

\subsubsection*{Support}
 
The distribution $p$ has support in the positive reals if and only if $\phi$ is analytic in the upper half of the complex plane and $\phi(t)$ grows at most polynomially as $\im t\to\infty$. This is theorem IX.16 in \cite{reed1975ii}, and can be seen as follows.

Suppose first that $p$ has support in the positive reals, so that we can write
\begin{equation}\label{phippositive}
    \phi(t)=\int_{0}^\infty dVe^{itV}p(V).
\end{equation}
If we allow $t$ to have a positive imaginary part, the integral will still be convergent, and differentiating with respect to $t$ will not spoil the convergence because of the exponential damping. Thus, $\phi$ is analytic in the upper half-plane. Moreover, in the limit $\im t\to\infty$ the exponential $e^{itV}$ vanishes for positive $V$, so $\phi$ will vanish unless $p(V)$ contains terms proportional to $\delta(V)$ or its derivatives. In a temperate distribution there can only be a finite number of terms like this. Taking into account that the Fourier transform of $\delta^{(n)}$ is proportional to $t^n$, this means that $\phi$ can grow at most polynomially in this limit.

Conversely, if $\phi$ is analytic in the upper half-plane and $\phi(t)$ grows at most polynomially as $\im t\to\infty$, then for $V<0$ we have
\begin{equation}
    p(V)=\frac{1}{2\pi}\int_{-\infty}^\infty dt\, e^{-itV}\phi(t)=\frac{1}{2\pi}\ointctrclockwise dt\, e^{-itV}\phi(t)=0,
\end{equation}
where in the second step we used the asymptotic property to close the contour in the upper half-plane, and in the third we used analyticity. Hence, $p$ has support in the positive reals.

Combining this result with (\ref{characteristic}) we see that the support condition is equivalent to saying that $Z(\Lambda)$ extends to an analytic function $Z(w)$ in the right half-plane $\re w>0$ which grows at most polynomially as $\re w\to\infty$. 

\subsubsection*{Positivity revisited}

Suppose that $p$ has support in the positive reals, and consider the function $\varphi(s)=\phi(is)$ with $s>0$, which is the Laplace transform of $p$,
\begin{equation}
    \varphi(s)=\int_{0}^\infty dV e^{-sV}p(V).
\end{equation}
If $p$ is positive, then $\varphi$ is what is called a completely monotonic function, i.e., $\varphi\ge 0$, $\varphi'\le 0$, $\varphi''\ge 0$, and so on; more compactly, $(-1)^n\varphi^{(n)}\ge 0$. In fact, the converse statement is also true: if $\varphi$ is completely monotonic, then $p$ is positive. This is Bernstein's theorem (theorem 1.4 in \cite{schilling2012bernstein}).

Thus, assuming that the support condition is satisfied, the positivity condition is equivalent to saying that $Z(\Lambda)$ is proportional (with a constant coefficient) to a completely monotonic function.

In order to use this result it is convenient to say some words about completely monotonic functions. Note first that a linear combination with positive coefficients of two completely monotonic functions is completely monotonic. Using Bernstein's theorem one easily shows that the product is also completely monotonic. These two properties together imply that the exponential of a completely monotonic function is completely monotonic. Simple examples of completely monotonic functions are $f(s)=e^{-\alpha s}$ and $f(s)=s^{-\alpha}$, with $\alpha\ge 0$.

\section{Statistics of the volume in various approximations}\label{sec:examples}

In this section, we will compute and study the probability distribution (\ref{prob}) for the total Euclidean volume using known results and proposals for the gravitational path integral in several different levels of approximation. In subsection \ref{sec:saddle} we begin with the saddle-point approximation, which gives explicit results in generic dimension $d$. In subsection \ref{sec:1loop} we study the one-loop corrections to the saddle-point result, focusing on dimensions $3$ and $4$. In $d=3$ there is an all-loop and a non-perturbative proposal for the gravitational path integral, which we study in subsection \ref{sec:all-loop}. Finally, subsection \ref{sec:liouville} is devoted to the case $d=2$. In this case the vacuum Einstein's equations with positive cosmological constant have no solutions, but, after coupling the theory to conformal matter, one obtains a model which has the 2-sphere as a classical solution. The dynamics of the geometry in this model is governed by the so-called timelike Liouville theory. In this context, the dependence of the gravitational path integral on the cosmological constant can be computed exactly, and from it one obtains an exact expression for the probability distribution.

\subsection{Saddle-point approximation ($d$ dimensions)}\label{sec:saddle}

We start with the computation of the probability distribution in the case where we use the saddle-point evaluation of $Z$. This is given by \eqref{Z_saddle}, which can be expressed in terms of the cosmological constant using \eqref{global} as
\begin{align}\label{Z_saddle_d-dim}
    Z(\Lambda)=\exp\left(\frac{C_d}{G\Lambda^{d/2-1}}\right)\qquad C_d=\frac{\Omega_{d-2}}{4}\left[\frac{(d-2)(d-1)}{2}\right]^{d/2-1}.
\end{align}
Note that this formula satisfies all the conditions of section \ref{sec:conditions}: it extends to an analytic function $Z(w)$ in the right half-plane (and beyond) which tends to $1$ as $\re w\to\infty$, and it is completely monotonic because it is the exponential of a completely monotonic function. Therefore, the distribution (\ref{prob}) will be positive and supported in the positive reals, as it should be. Note also that, from the behavior as $\re w\to\infty$, we can anticipate that $p(V)$ will have a term proportional to $\delta(V)$ (see the discussion under (\ref{phippositive})).

Before computing the probability distribution, we can quickly obtain the expectation value and the variance of the volume from (\ref{variance}),
\begin{align}\label{varsaddle}
\langle \mathcal{V}\rangle = \frac{4\pi(d-2)C_d}{\Lambda^{d/2}} =\Omega_{d}\ell^{d} = V_{cl}\qquad\var(\mathcal{V})=d\frac{4\pi G}{\Lambda}V_{cl},
\end{align}
where $V_{cl}$ denotes the classical volume, namely the volume of the sphere or Euclidean dS. In the limit $G\to 0$ the variance vanishes, and hence the probability distribution is a delta function centered at the classical volume, as it should be.
Note however that, away from this limit, the distribution will have some width, even though we are working in the saddle-point approximation.

\begin{figure}
    \centering
    \includegraphics[width=1\linewidth]{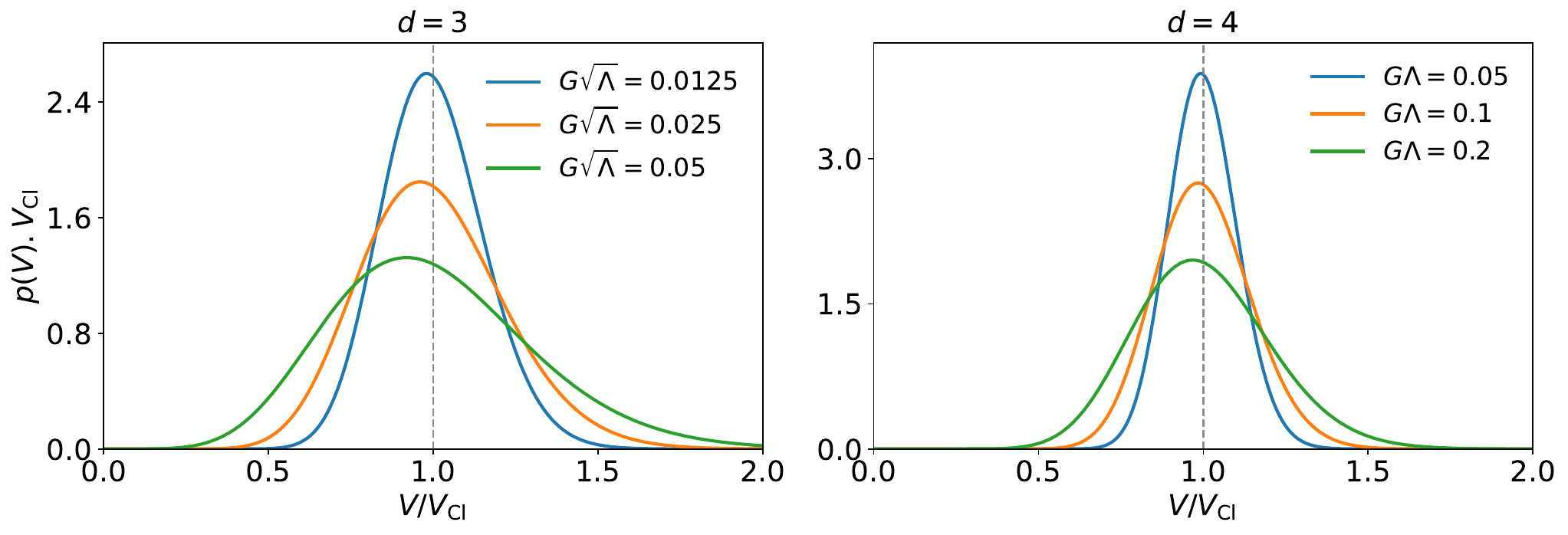}
    \caption{Probability distribution for the volume obtained via the saddle point approximation to the gravitational path integral for $d=3, 4$ and various values of $G\Lambda^{d/2-1}$. In our universe ($d=4$, $G\Lambda\sim 10^{-122}$), the distribution is extremely concentrated around the classical value.}
    \label{fig:saddle}
\end{figure}

Now, let us proceed with the probability distribution. From (\ref{prob}) and (\ref{Z_saddle_d-dim}) we have
\begin{align}
    p(V) = \frac{1}{2\pi Z(\Lambda)}\int_{-\infty}^{\infty}dt\,e^{-itV}\exp\left[ \frac{C_d}{G(\Lambda-it8\pi G)^{d/2-1}}\right].
\end{align}
The Fourier transform above can be computed by Taylor-expanding the exponential and using the known result
\begin{align}\label{Fourier}
    \frac{1}{2\pi}\int_{-\infty}^{\infty}dt\,\frac{e^{-itx}}{(a-it)^{\beta}} = \frac{1}{\Gamma(\beta)} \Theta(x) x^{\beta-1}e^{-ax}
    \qquad\text{for }\beta>0,
\end{align}
where $\Theta$ is the step function. Doing this, we obtain
\begin{align}\label{tree level p}
    p(V)=\frac{1}{Z(\Lambda)}\left[\delta(V) + \frac{\Theta(V)}{V} \exp\left({-\frac{\Lambda V}{8\pi G}} \right)W_{d/2-1,0}\left(\frac{8\pi C_d V^{d/2-1}}{(8\pi G)^{d/2}}\right)\right],
\end{align}
where $W_{\lambda,\mu}$ is the Wright function,
\begin{align}\label{Wright}
    W_{\lambda,\mu}(x) =\sum_{k=0}^{\infty}\frac{x^k}{k!\Gamma(\lambda k+\mu)}.
\end{align}
Note that, for $G\Lambda^{d/2-1}\ll 1$ (the regime where the saddle-point approximation is justified), the partition function is exponentially large and therefore the term proportional to $\delta(V)$ in (\ref{tree level p}) is negligible (because it integrates to an exponentially small number, whereas $p$ integrates to 1). We plot this distribution for $d=3,4$ and for various values of $G\Lambda^{d/2-1}$ in \figref{fig:saddle}. We see that, as $G\Lambda^{d/2-1}$ becomes smaller, the distribution concentrates around the classical volume, in agreement with the discussion above.

For volumes much larger than the Planck volume, $V\gg G^{d/(d-2)}$, the probability can be written in a simpler form using the asymptotic formula for the Wright function \cite{wright1935asymptotic}
\begin{equation}\label{wrightasym}
    W_{\lambda,\mu}(x)\approx \frac{1}{\sqrt{2\pi(\lambda+1)}}(\lambda x)^{\frac{1-2\mu}{2(\lambda+1)}}\exp\left[\frac{\lambda+1}{\lambda}(\lambda x)^{\frac{1}{\lambda+1}}\right]\qquad x\gg 1.
\end{equation}
Substituting into (\ref{tree level p}), and using (\ref{varsaddle}) to express $C_d$ in terms of $V_{cl}$, we obtain
\begin{align}
\begin{aligned}
    &p(V)\approx\sqrt{\frac{\Lambda}{8\pi^2 dG}}V_{cl}^{1/d}V^{-1/2-1/d}\exp\left[-\frac{\Lambda}{8\pi G}f(V)\right]\\
    &f(V)=V-\frac{d}{d-2}V_{cl}^{2/d} V^{1-2/d}+\frac{2}{d-2}V_{cl}.
\end{aligned}
\end{align}
Note that $f(V)\ge 0$, with the equality holding only at $V=V_{cl}$. This gives another way to see that, in the limit $G\to 0$, the distribution tends to a delta function centered at the classical volume.

\subsection{One-loop approximation ($d=3,4$)}\label{sec:1loop}

In the cases $d=3,4$ there are one-loop computations of the gravitational path integral in pure gravity \cite{anninos_2022_quantum}{\footnote{There are also explicit one-loop results in some other dimensions, but we concentrate on these two.}}. In $d=4$, the proposed result in terms of the cosmological constant is
\begin{align}\label{Z1loop4d}
    \log Z(\Lambda) = \frac{3\pi}{G\Lambda }-5\log\frac{3\pi}{G\Lambda}-\frac{571}{45}\log\left(\sqrt{\frac{3}{\Lambda}}\frac{1}{\ell_{\text{ren}}}\right)+\mathcal{O}(1),
\end{align}
where the $\mathcal{O}(1)$ terms are independent of $\Lambda$ (and include the phase factors), and $\ell_{\text{ren}}$ is a renormalization scale. The first term above is the saddle-point contribution, see (\ref{varsaddle}). This formula for $Z$ satisfies the support condition of section \ref{sec:conditions}, but not the positivity condition, because the one-loop corrections break complete monotonicity. However, this happens only when these corrections become comparable to the saddle-point contribution, i.e., outside the domain of validity of the one-loop approximation, so it is nothing to worry about.

Substituting (\ref{Z1loop4d}) into (\ref{variance}) we obtain
\begin{align}\label{Var dS4}
\langle \mathcal{V}\rangle =V_{cl}-\frac{1021}{90}\frac{8\pi G}{\Lambda}\qquad\var(\mathcal{V})= \frac{16\pi G}{\Lambda}V_{cl}-\frac{1021}{90}\left(\frac{8\pi G}{\Lambda}\right)^2,
\end{align}
with $V_{cl}=24\pi^2/\Lambda^2$.
The first term in each expression is the saddle-point contribution, and the second is the one-loop correction. We see that the quantum correction makes the expectation value and the variance smaller. In fact, this illustrates the discussion above: for sufficiently large values of $G$, the expectation value and the variance become negative, signaling a breakdown of positivity of the probability distribution, but of course in this regime we cannot trust the one-loop approximation.

\begin{figure}
    \centering
    \includegraphics[width=1\linewidth]{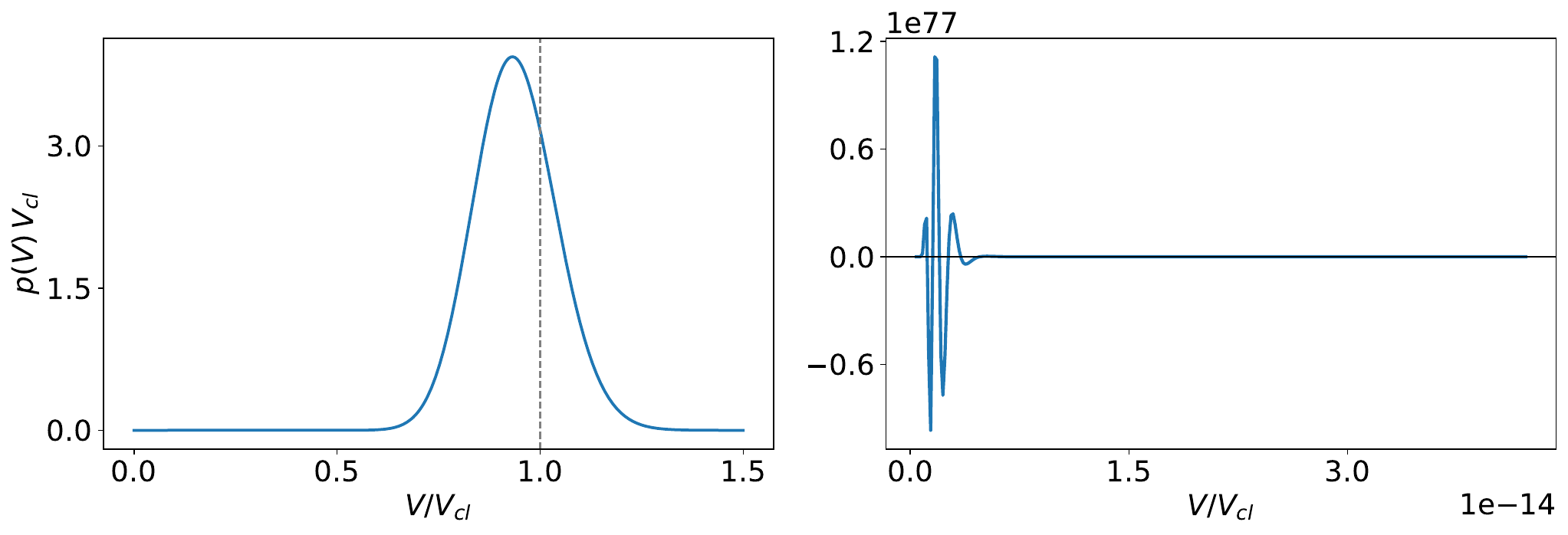}
    \caption{Left panel: one-loop probability distribution for the volume in 4 dimensions, Eqs.~(\ref{1loop1}) and (\ref{1loop2}), with $G\Lambda=0.05$. Right panel: a zoom of the same plot near the origin, using the regularization $\theta_\epsilon(V)=e^{-\epsilon/V}$ for the step function at $V>0$, with $\epsilon/V_{cl}\approx 4\times 10^{-14}$. We see that, very close to the origin, $p$ takes negative values. As the regulator $\epsilon$ is sent to 0, the negativities concentrate at the origin and become distributional.}
    \label{fig:negativities}
\end{figure}

\begin{figure}
    \centering
    \includegraphics[width=0.7\linewidth]{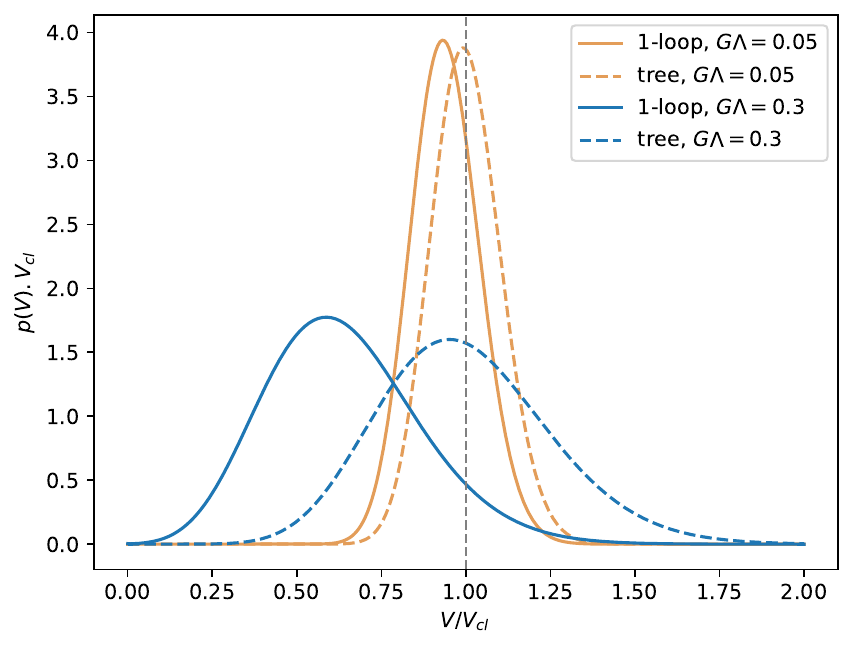}
    \caption{Probability distribution $p(V)$ obtained via the saddle-point and one-loop approximations to the gravitational path integral, in 4 dimensions and for two different values of $G\Lambda$.}
    \label{fig:1l_dS4}
\end{figure}

Let us compute the probability distribution (\ref{prob}) in this case. We have
\begin{align}\label{integral ds4}
    p(V)=\frac{1}{2\pi}\exp\left(-\frac{3\pi}{G\Lambda}\right)\int_{-\infty}^{\infty}dt\,e^{-itV}\exp\left[\frac{3\pi}{G(\Lambda-it8\pi G)}\right]\left(1-\frac{it8\pi G}{\Lambda}\right)^{1021/90}.
\end{align}
Since integer powers of $-it$ correspond to derivatives under the Fourier transform, we can rewrite this equation as
\begin{equation}\label{1loop1}
    p(V)=\left(1+\frac{8\pi G}{\Lambda}\frac{d}{dV}\right)^{12}q(V),
\end{equation}
where
\begin{equation}
    q(V)=\frac{1}{2\pi}\exp\left(-\frac{3\pi}{G\Lambda}\right)\int_{-\infty}^\infty dt\,e^{-itV}\exp\left[\frac{3\pi}{G(\Lambda-it8\pi G)}\right]\left(1-\frac{it8\pi G}{\Lambda}\right)^{-\beta}
\end{equation}
and $\beta=12-1021/90=59/90$.
The latter Fourier transform can be computed by the same method as in the previous subsection, giving
\begin{equation}\label{1loop2}
    q(V)=\left(\frac{\Lambda}{8\pi G}\right)^{\beta}\Theta(V)V^{\beta-1}\exp\left(-\frac{3\pi}{G\Lambda}-\frac{\Lambda V}{8\pi G}\right)W_{1,\beta}\left(\frac{3V}{8G^2}\right).
\end{equation}
Eqs.~(\ref{1loop1}) and (\ref{1loop2}) give the probability distribution for the volume in the one-loop approximation, and in the case $d=4$. For $V\gg G^2$ we can approximate
\begin{equation}
    q(V)\approx\sqrt{\frac{\Lambda}{32\pi^2G}}V_{cl}^{1/4-\beta/2}V^{\beta/2-3/4}\exp\left[-\frac{\Lambda}{8\pi G}\left(\sqrt{V}-\sqrt{V_{cl}}\right)^2\right],
\end{equation}
where we have used the asymptotic formula (\ref{wrightasym}). Note that $q$ is positive (in fact it is also normalized, because its Fourier transform is $1$ at the origin). The negativities of $p$ associated with the lack of complete monotonicity of $Z$ are due to the derivatives in (\ref{1loop1}). We have numerical evidence that these negativities are distributional and only occur at the origin, as we show in 
\figref{fig:negativities}. In \figref{fig:1l_dS4} we compare the one-loop distribution to the tree-level one for various values of $G\Lambda$.

The case $d=3$ is analogous. The one-loop partition function is given by \cite{anninos_2022_quantum}
\begin{align}\label{Z1loop3d}
    \log Z(\Lambda) = \frac{\pi}{2G\sqrt{\Lambda}}- 3\log \frac{\pi}{2G\sqrt{\Lambda}} + \mathcal{O}(1),
\end{align}
the first term being the saddle point contribution of (\ref{Z_saddle_d-dim}). The second term breaks the complete monotonicity of $Z$ (and hence the positivity of the associated probability distribution), although only outside the regime where the approximation is valid. The corresponding expectation value and variance of the volume are
\begin{align}\label{Exp value_one_loop_ds3}
\langle \mathcal{V}\rangle = V_{cl}-\frac{3}{2}\frac{8\pi G}{\Lambda}\qquad\var(\mathcal{V})= \frac{12\pi G}{\Lambda}V_{cl}-\frac{3}{2}\left(\frac{8\pi G}{\Lambda}\right)^2,
\end{align}
with $V_{cl}=2\pi^2/\Lambda^{3/2}$. The probability distribution for the volume is
\begin{align}\label{prob1loop3d}
    \begin{aligned}
        &p(V)=\left(1+\frac{8\pi G}{\Lambda}\frac{d}{dV}\right )^2q(V)\\
        &q(V)=\Theta(V)\sqrt{\frac{\Lambda}{8\pi GV}}\exp\left(-\frac{\pi}{2G\sqrt{\Lambda}}-\frac{\Lambda V}{8\pi G}\right)W_{1/2,1/2}\left(\sqrt{\frac{\pi V}{32G^3}}\right),
    \end{aligned}
\end{align}
and for $V\gg G^3$ we can approximate
\begin{equation}
    q(V)\approx \sqrt\frac{\Lambda}{24\pi^2 GV}\exp\left[-\frac{\Lambda}{8\pi G}\left(V-3V_{cl}^{2/3}V^{1/3}+2V_{cl}\right)\right].
\end{equation}

\subsection{All-loop and non-perturbative proposals ($d=3$)}\label{sec:all-loop} 

In \cite{anninos_2022_quantum} the authors propose an all-loop result for the Euclidean gravitational path integral around the 3-sphere
by working with the Chern-Simons formulation of 3D gravity. Their partition function is 
(see equation (H.24) in \cite{anninos_2022_quantum}\footnote{We set $l=0$ in their expression, which corresponds to standard Einstein gravity.})
\begin{align}\label{propuesta Anninos}
    Z=\frac{2e^{2\pi\kappa}}{\sqrt{4+\kappa^2}}\sin\left(\frac{\pi}{2+i\kappa}\right)\sin\left(\frac{\pi}{2-i\kappa}\right)\qquad \kappa=\frac{1}{4G\sqrt{\Lambda}}.
\end{align}
This partition function reproduces the one-loop result (\ref{Z1loop3d}) at small values of $G\sqrt{\Lambda}$. Moreover, remarkably, it satisfies all the conditions of section \ref{sec:conditions} (we have checked complete monotonicity numerically), so the corresponding probability distribution will be positive and supported in the positive reals. In particular, the higher-loop contributions correct the negativities of the one-loop approximation. A peculiarity of this formula is that it gives $\log Z\approx 2\pi\kappa$ both for large and small values of $\kappa$, which means that, in the highly quantum regime $G\sqrt{\Lambda}\gg 1$, all moments of ${\mathcal V}$ are well approximated by the saddle-point expressions.

There is also a different proposal for the 3D sphere partition function given in \cite{Collier:2025lux}. In the cited work the authors propose that the  non-perturbative expression for the gravitational path integral in 3 dimensions with $\Lambda >0$ is given by (for $|b|<1$)
\begin{align}\label{CLS proposal}
    Z = -b^2\left[\frac{4 \sin(\pi b^2) \sin (\pi b^{-2})}{(1-b^4)}\right]^2
\end{align}
Where $b^2$ is purely imaginary. Although the relation between $b^2$ and the gravitational parameters is not entirely clear\footnote{See comment below (5.8) in \cite{Collier:2025lux}}, one can make a proposal for this relation by demanding that the imaginary part of the central charge of the complex Liouville theory matches with the analytic continuation of the Brown-Henneaux formula. That is,
\begin{align}
    b^2+b^{-2}=\frac{i}{4G\sqrt{\Lambda}}=i\kappa.
\end{align}
By demanding this, there are two possible values for $b^2$,
\begin{align}
    b^2=i\left(\frac{\kappa}{2}\pm\sqrt{\frac{\kappa^2}{4}+1}\right).
\end{align}
Keeping the minus branch since it is the one that satisfies $|b|<1$, it is easy to see that the formula (\ref{CLS proposal}) for the partition function reproduces the one-loop result (\ref{Z1loop3d}) for $G\sqrt{\Lambda}\ll 1$. Moreover, with this choice, the formula satisfies all the conditions of section \ref{sec:conditions}, like the all-loop proposal. As in that case, here we checked complete monotonicity numerically. In the quantum regime $G\sqrt{\Lambda}\gg 1$, Eq.~(\ref{CLS proposal}) gives $\log Z\approx c+\kappa/2$, where $c$ is a constant, and therefore the expectation value and the variance of ${\mathcal V}$ are well approximated by the saddle-point expressions divided by $4\pi$.

\begin{figure}
    \centering
    \includegraphics[width=1\linewidth]{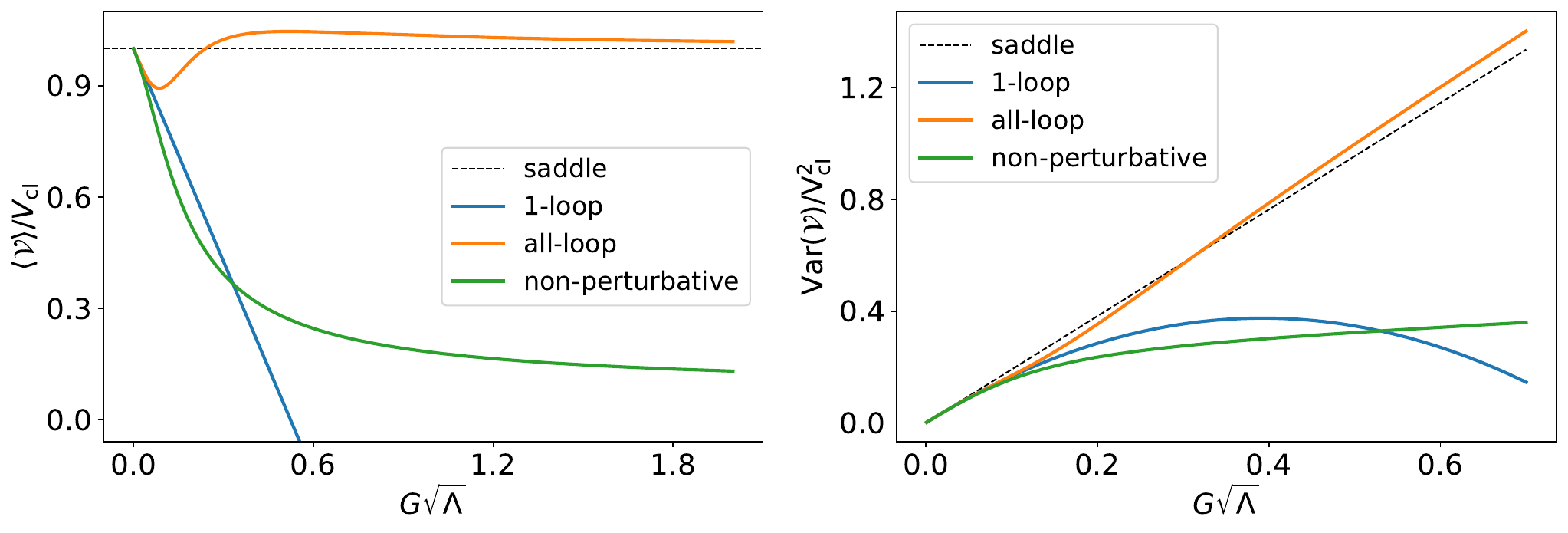}
    \caption{Expectation value and variance of the volume in 3 dimensions, as a function of $G\sqrt{\Lambda}$, according to (i) the saddle-point approximation, (ii) the one-loop approximation, (iii) the all-loop proposal for the partition function given in \cite{anninos_2022_quantum}, and (iv) the non-perturbative proposal of \cite{Collier:2025lux}.}
    \label{fig:var3d}
\end{figure}

In \figref{fig:var3d} we plot the expectation value and the variance of ${\mathcal V}$ as a function of $G$ according to the all-loop proposal (\ref{propuesta Anninos}) and the non-perturbative proposal (\ref{CLS proposal}), in comparison with the saddle-point and one-loop approximations. We see that, according to both proposals, the expectation value and the variance remain positive at all values of $G$, correcting the negativities of the one-loop result, in consonance with the discussion above. The plot is also consistent with the behavior discussed above at large values of $G\sqrt{\Lambda}$. In fact, we see that the saddle-point results give a reasonably good approximation to the all-loop expressions not only at large and small values of $G\sqrt{\Lambda}$, but also for intermediate values.

\begin{figure}
    \centering
    \includegraphics[width=1\linewidth]{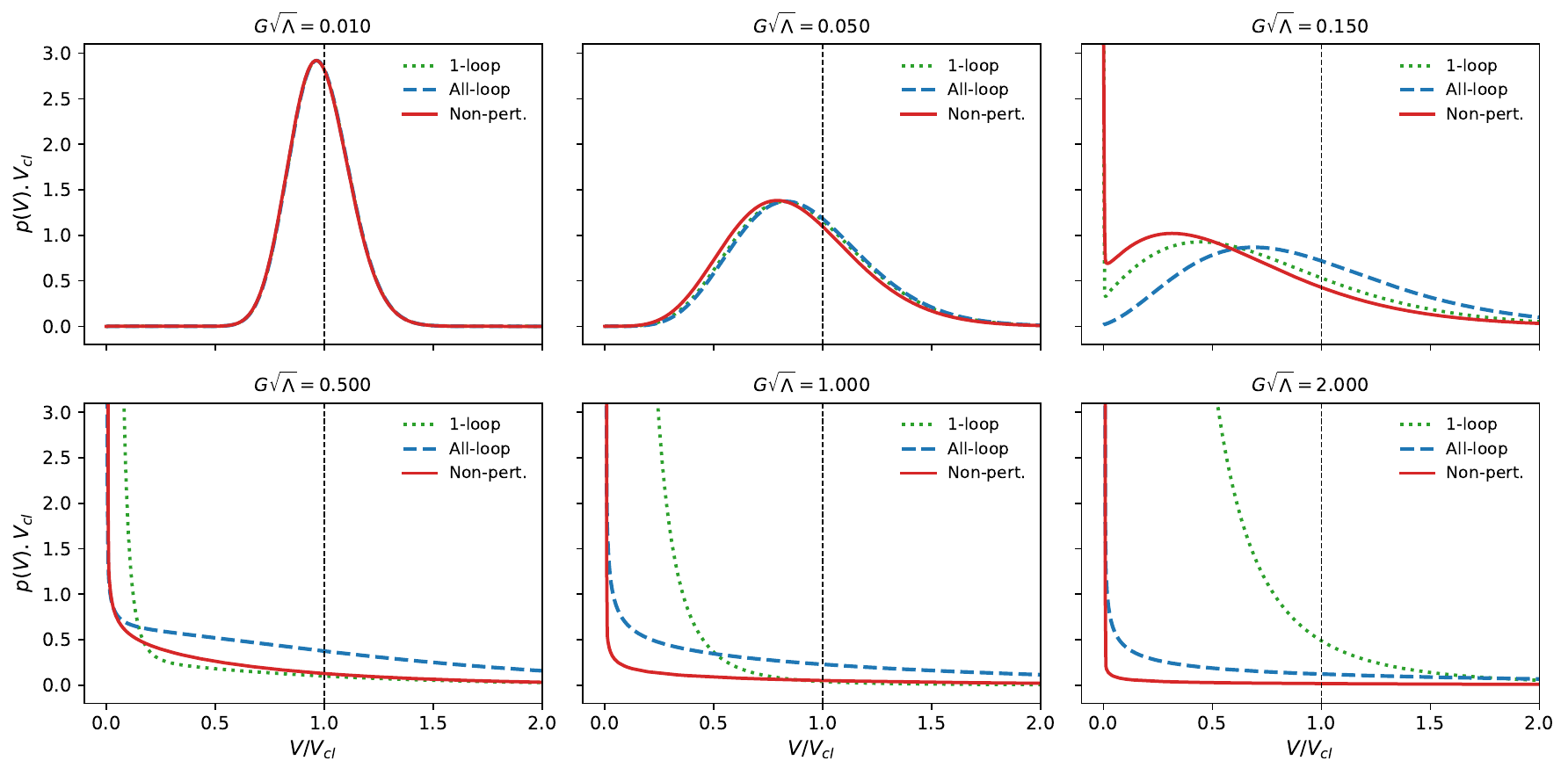}
    \caption{Probability distribution $p(V)$ in 3 dimensions, obtained via the all-loop and non-perturbative proposals for the partition function and via the one-loop approximation. We plot these distributions for various values of $G\sqrt{\Lambda}$.}
    \label{fig:prob3d}
\end{figure}

We compute numerically the probability distribution for the volume associated with both proposals for the partition function, and plot it
for different values of $G\sqrt{\Lambda}$ in \figref{fig:prob3d}, in comparison with the one-loop probability distribution obtained in the previous subsection, Eq.~(\ref{prob1loop3d}). We see that, for $G\sqrt{\Lambda}< 0.1$ the three distributions essentially coincide. For $G\sqrt{\Lambda}>0.1$ we start to see quantitative disagreements, although the all-loop and non-perturbative distributions remain qualitatively similar. Both distributions tend to concentrate around increasingly small volumes, leaving a tail responsible for the fact that the expectation value and the variance of the volume do not go to zero as $G\sqrt{\Lambda}$ grows. At large values of $G\sqrt{\Lambda}$, the one-loop distribution (which is not to be trusted in that regime) broadens, but this broadening is compensated by the
distributional negativities of $p$ at the origin, which give rise to the negative values of $\langle{\mathcal V}\rangle$ and $\var({\mathcal V})$ found in that case (and also guarantee that $p$ remains normalized despite appearances).

We close this section with a brief digression. It may sound surprising that the dS partition function in 3 dimensions is not one-loop exact, unlike what happens in AdS \cite{Maloney:2007ud} and asymptotically flat gravity \cite{Witten:1989sx,Leston:2023ugd}. We note, however, that, in the AdS case, one-loop exactness relies on a specific choice of boundary conditions, namely those corresponding to the thermal partition function, i.e., that the geometry should have a torus as conformal boundary. Changing the boundary conditions may spoil one-loop exactness, as happens for example when one replaces the torus by a double torus \cite{Headrick:2015gba}. From this perspective, it is a bit less surprising that the dS result is not one-loop exact, given that the boundary conditions are different (the geometry in this case has no conformal boundary). Moreover, in the limit $\Lambda\to 0$, after removing divergent terms, the logarithm of (\ref{propuesta Anninos}) and (\ref{CLS proposal}) becomes of order $G^0$, i.e., one-loop exact.

\subsection{An exact result ($d=2$)}\label{sec:liouville}

The case $d=2$ is special because of the Gauss-Bonnet theorem, 
\begin{equation}
    \int d^2x\sqrt{g}R=8\pi(1-h),
\end{equation}
where $h$ is the genus (i.e., the number of handles) of the manifold.
Thus, the integral of the Ricci scalar in 2 dimensions is a topological invariant, which implies that 
the Einstein tensor vanishes identically and hence the vacuum Einstein's equations with non-zero cosmological constant have no solutions.

To remedy this situation, we can couple the theory to matter. Writing explicitly the sum over topologies implicit in (\ref{Z}), the partition function is
\begin{equation}
    Z=\sum_h\exp\left(\frac{1-h}{2G}\right)Z_h\qquad Z_h=\int{\mathcal D}g \exp\left(-\frac{\Lambda}{8\pi G}\int d^2x\sqrt{g}\right)Z_m[g],
\end{equation}
where the sum is over values of the genus, and $Z_m$ is the matter partition function. Let us assume that the matter is conformal, and let us work in the Weyl gauge, $g_{\mu\nu}=e^\sigma\tilde g_{\mu\nu}$, with $\tilde g$ a fixed fiducial metric. The dependence of $Z_m$ on $\sigma$ is dictated by the Weyl anomaly, and, under the Distler-Kawai hypothesis \cite{distler_1989_conformal}, it leads to
an expression of $Z_h$ as the product of three CFT partition functions evaluated on the fiducial background,
\begin{equation}
    Z_h=Z_LZ_m Z_{gh}.
\end{equation}
Here, $Z_L$ is the partition function of Liouville theory \cite{polyakov_1981_quantum}, and $Z_{gh}$ that of the Faddeev-Popov $\mathbf{bc}$-ghosts needed to properly fix the Weyl gauge.
To ensure background independence, the central charges of the Liouville and matter theories have to satisfy the consistency condition 
\begin{align}\label{central charge balance}
    c_L + c_m-26 =0 ,
\end{align}
where the $-26$ piece is the central charge of the ghost system. If we consider matter theories with $c_m\ge 25$, we need $c_L\le 1$, which forces us to deal with the so-called timelike Liouville theory. The corresponding partition function is
\begin{align}\label{liouville}
    \begin{aligned}
        &Z_L=\int{\mathcal D}\phi\, e^{-S_L[\phi]}\\
        &S_L[\phi] = \frac{1}{4\pi}\int d^2x \sqrt{\tilde{g}}\left(-\tilde{g}^{\mu\nu}\partial_{\mu}\phi \partial_{\nu} \phi + Q\tilde{R}\phi + 4\pi \mu e^{2b\phi}\right),
    \end{aligned}
\end{align}
where $b,\mu>0$ and $Q=b-1/b$. The central charge of timelike Liouville is $c_L = 1-6Q^2$, which, together with (\ref{central charge balance}), implies
\begin{equation}\label{cmb}
    c_m=25+6Q^2=25+6\left(b-\frac{1}{b}\right)^2.
\end{equation}
The Liouville field $\phi$ is related to the Weyl mode $\sigma$ by $\sigma=2b\phi$. The constant $\mu$ is called ``cosmological constant'', because the exponential term in the timelike Liouville action comes from the cosmological constant term in the Einstein-Hilbert action, and $\mu$ comes from a renormalization (and a redefinition) of $\Lambda$ \cite{zamolodchikov_lectures,ginsparg_2024_lectures}. A feature of the timelike Liouville action is that the kinetic term has the ``wrong sign'', making the action unbounded from below. This is the 2-dimensional analog of the conformal mode problem encountered in higher-dimensional gravity \cite{Polchinski:1988ua,dionysiosanninos_2021_the}.

The classical limit of timelike Liouville is obtained by letting $b\to 0$ while keeping $\sigma=2b\phi$ and $M\equiv 2\pi\mu b^2$ fixed \cite{harlow_2011_analytic}. Indeed, in this limit $S_L$ grows as $1/b^2$, so $Z_L$ can be computed in the saddle-point approximation. The equation of motion is
\begin{equation}
    e^{-\sigma}(\tilde R-\tilde\Delta\sigma)=4M.
\end{equation}
This equation is the statement that the physical metric $g$ has constant curvature $R=4M$, so the saddle point is the sphere of radius
$\ell=1/\sqrt{2M}$. Thus,
we ended up with a 2D theory of gravity coupled with conformal matter that has (Euclidean) $dS_2$ as a classical solution.\footnote{Instead of $c_m\ge 25$, another consistent choice is $c_m\le 1$, in which case the Liouville theory is the spacelike one, which has $c_L\ge 25$. The classical limit of spacelike Liouville describes surfaces of constant negative curvature, so in order to have a model of $dS_2$ it is crucial to have $c_m \geq 25$.} Note from (\ref{cmb}) that $c_m\to\infty$ in the classical limit.

Let us now study the statistics of the volume in the context of this model. In fact, since we are working in 2 dimensions, instead of the volume we should speak of the area  
${\mathcal A}[\phi]=\int d^2x\sqrt{g}=\int d^2x\sqrt{\tilde g}\, e^{2b\phi}$. Note from (\ref{liouville}) that $\mu$ plays exactly the same role as $\Lambda/8\pi G$, so
the expectation value and the variance are
\begin{equation}\label{variance2}
    \langle{\mathcal A}\rangle=-\partial_\mu\log Z\qquad\var({\mathcal A})=\partial_{\mu}^2\log Z,
\end{equation}
and the probability distribution is
\begin{equation}\label{prob2d}
    p(A)=\frac{1}{2\pi}\int_{-\infty}^{\infty}dt\,e^{-itA} \frac{Z(\mu-it)}{Z(\mu)}.
\end{equation}
The consistency conditions that $Z$ must satisfy in order for $p$ to be positive and supported in the positive reals are those of section \ref{sec:conditions} with $\Lambda$ replaced by $\mu$.

The crucial advantage of this model with respect to the higher-dimensional ones considered above is that the $\mu$-dependence of $Z$ can be studied exactly. Indeed, the Liouville actions with two different cosmological constants are related by
\begin{equation}
    S_{L}^{(\mu)}[\phi]=S_{L}^{(\mu_0)}[\phi']-\frac{Q(1-h)}{b}\log(\mu/\mu_0)\qquad \phi'=\phi+\frac{1}{2b}\log(\mu/\mu_0),
\end{equation}
where we used the Gauss-Bonnet theorem, and therefore
\begin{equation}\label{kpz}
    Z_L(\mu)=\left(\frac{\mu}{\mu_0}\right)^{Q(1-h)/b}Z_L(\mu_0).
\end{equation}
This result is known as KPZ scaling \cite{KPZ}. Since the matter and ghost partition functions do not depend on $\mu$, for the full partition function we have
\begin{equation}\label{fullkpz}
    Z(\mu)=\sum_h e^{(1-h)/2G}\left(\frac{\mu}{\mu_0}\right)^{Q(1-h)/b}Z_h(\mu_0).
\end{equation}
Note that the only role of Newton's constant in this context is to control the weight of the different topologies. Taking the limit $G\to 0$, so that only the sphere topology ($h=0$) contributes, we obtain a very simple $\mu$-dependence,
\begin{equation}\label{Zliouville}
    Z=\alpha\mu^{1-1/b^2},
\end{equation}
where $\alpha$ is $\mu$-independent. This partition function satisfies the support condition of section \ref{sec:conditions} for any value of $b$; the positivity condition is also satisfied provided that $b\le 1$, so we restrict to this range (which covers all possible values of $c_m$).

\begin{figure}
    \centering
    \includegraphics[width=0.7\linewidth]{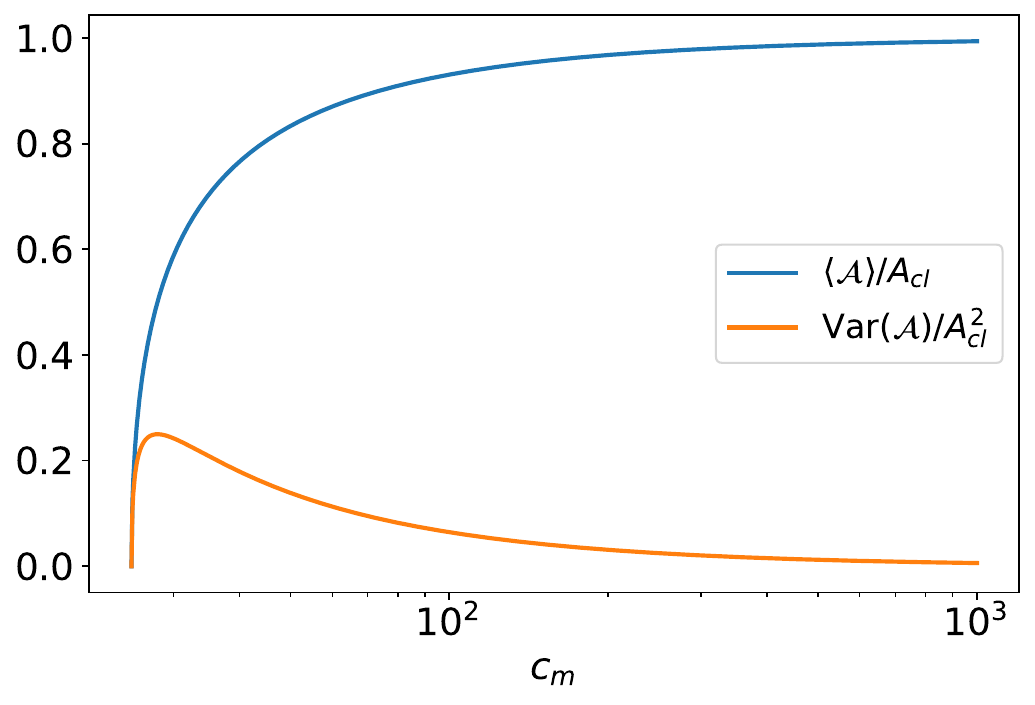}
    \caption{Expectation value and variance of the area (in 2 dimensions) as a function of the matter central charge $c_m$. The classical limit is $c_m\to\infty$, and at the opposite end we have $c_m=25$.}
    \label{fig:Area}
\end{figure}

Substituting (\ref{Zliouville}) into (\ref{variance2}) we obtain the expectation value and variance of the area, which can be written in terms of the classical area
$A_{cl}=4\pi\ell^2=2\pi/M=1/\mu b^2$ as
\begin{equation}
    \langle{\mathcal A}\rangle=A_{cl}(1-b^2)\qquad \var({\mathcal A})=A_{cl}^2b^2(1-b^2).
\end{equation}
In the classical limit, $b\to 0$, the expectation value is the classical area and the variance vanishes; in other words, the probability distribution is a delta function centered at the classical area, as expected. More surprisingly, both the expectation value and the variance vanish in the opposite limit, $b\to 1$ ($c_m\to25$), so in that limit the distribution is a delta function centered at the origin. 
We plot these results in terms of the matter central charge in \figref{fig:Area}.

\begin{figure}
    \centering
    \includegraphics[width=0.7\linewidth]{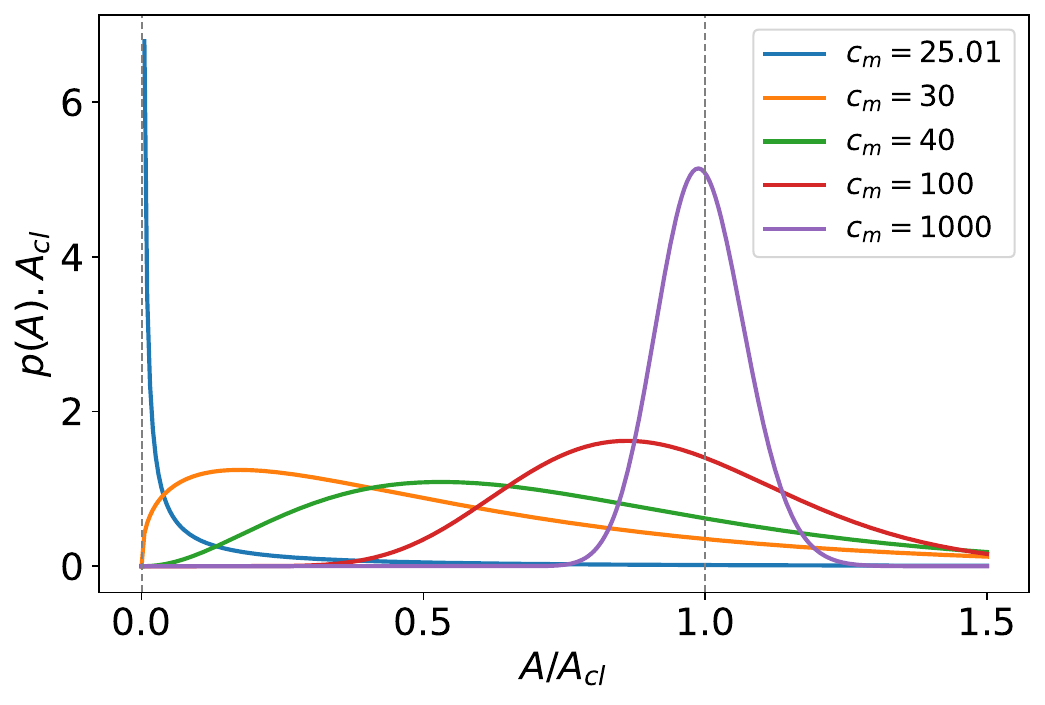}
    \caption{Probability distribution for the area of the 2D Euclidean manifold for various values of $c_m$.}
    \label{fig:P(A)}
\end{figure}

The probability distribution is also easily computed using (\ref{Fourier}),
\begin{align}\label{probliouville}
    p(A)&=\frac{1}{2\pi}\int_{-\infty}^\infty dt\, e^{-itA}\left(1-\frac{it}{\mu}\right)^{1-1/b^2}\nonumber\\
    &=\frac{\mu^{1/b^2-1}}{\Gamma(1/b^2-1)}\Theta(A)A^{1/b^2-2}e^{-\mu A}\nonumber\\
    &=\frac{1}{\Gamma(1/b^2-1)(b^2A_{cl})^{1/b^2-1}}\Theta(A)A^{1/b^2-2}\exp\left(-\frac{A}{b^2A_{cl}}\right).
\end{align}
One can easily check that this distribution concentrates around the classical area for small $b$, and around the origin for $b$ close to 1. Hence, 
$p(A)$ approaches $\delta(A-A_{cl})$ when $b\to 0$ and $\delta(A)$ when $b\to 1$, in agreement with the discussion above. In \figref{fig:P(A)} we plot this distribution for various values of $c_m$.

We close this subsection by studying qualitatively the effects of the higher-order topologies. Suppose that $G$ is non-zero but small enough that $Z(\mu)$ is well approximated by the sphere contribution. To compute the probability distribution we need $Z(\mu-it)$, which according to (\ref{fullkpz}) can be written as
\begin{equation}\label{topologies}
    Z(\mu-it)=\sum_{h}\exp\left\{(1-h)\left[\frac{1}{2G}+\frac{Q}{b}\log\left(1-\frac{it}{\mu}\right)\right]\right\}Z_h(\mu).
\end{equation}
By hypothesis, at $t=0$ this sum is dominated by the sphere. It will remain so as long as the coefficient of $1-h$ has a large enough real part, which corresponds to
\begin{equation}
    t^2\ll\mu^2\left\{\exp\left[\frac{1}{G(b^{-2}-1)}\right]-1\right\}\equiv t_{c}^2.
\end{equation}
For $|t|\gtrsim t_{c}$, 
the contribution of higher-order topologies may be relevant. 
Hence, Eq.~(\ref{probliouville}) correctly captures the behavior of $p$ on scales much larger than $1/t_c$; on smaller scales there may be significant corrections from higher-order topologies.

The effects of higher-order topologies do not seem to alter the conclusion that $p(A)=\delta(A)$ when $b=1$. Indeed, in this case we have $Q=0$, and by (\ref{kpz}) the Liouville partition function is $\mu$-independent regardless of the topology. This makes the full partition function $\mu$-independent, see (\ref{fullkpz}), which in turn implies that $p$ is the Fourier transform of 1, i.e., the delta function centered at the origin. We take this argument with caution, however, because at the point $b=1$ there might be subtleties with KPZ scaling, and the Liouville partition function might be logarithmically dependent on $\mu$ instead of $\mu$-independent. If present , these subtleties do not play a role when only the sphere contributes, because in that case we have the probability distribution for generic values of $b$ and we can see how it behaves as $b$ approaches 1.

\section{Discussion and open questions}\label{discussion}

The Euclidean formulation of quantum gravity has naturally associated a formal probability distribution on Riemannian manifolds, namely $e^{-I}/Z$, where $I$ is the Euclidean action and $Z$ the partition function. In the context of dS quantum gravity, the statistics of the total volume (a diffeomorphism-invariant quantity) according to this distribution is encoded in the $\Lambda$-dependence of $Z$. In this paper we have used this observation, together with known results for the partition function of sphere topology, to obtain the probability distribution for the volume 
in several different levels of approximation: saddle point in $d$ dimensions, one loop in $d=3,4$, an all-loop and a non-perturbative proposal in $d=3$, and an exact result in $d=2$ (in the context of Liouville theory). In all cases we find a reasonable behavior: in the classical limit, the probability distribution is a delta function centered at the classical volume, and as quantum effects are turned on the distribution spreads. Moreover, the distribution is positive and supported in the positive reals, as it should (in the one-loop approximation there are negativities at the origin, but we have provided evidence that they are corrected by higher-loop effects). We note that in 4 dimensions there are results for the one-loop partition function around $\mathbb{CP}^2$ and $S^2\times S^2$ \cite{Anninos:2025ltd, Volkov:2000ih}, it would be interesting to obtain the probability distribution including these contributions and study how it gets corrected.

A common feature of all our results is that, near the classical limit, quantum effects make the expectation value of the volume smaller. One possible way to understand this is as follows. To compute the one-loop contribution to the partition function, one replaces the Euclidean action by the term $I_2$ quadratic in the metric perturbation, namely the departure from the metric of the sphere.
This quadratic action
is not bounded from below, but after an appropriate Wick rotation it becomes positive. Since it is proportional to $1/G$, this makes $e^{-I_2}$ an increasing function of $G$. Naively (i.e., ignoring that there are divergences that need to be renormalized), this implies that $Z_2$ (the one-loop contribution to the partition function) is itself proportional to an increasing function of $G$, with a constant coefficient (a phase coming from the Wick rotation). Moreover, $Z_2$ is dimensionless, so it has to depend on $G$ and $\Lambda$ through the combination $G\Lambda^{d/2-1}$. It follows that $Z_2$ is proportional (with a constant coefficient) to an increasing function of $\Lambda$, and hence the one-loop contribution to the expectation value of the volume, $\langle{\mathcal V}\rangle_2=-8\pi G\partial_\Lambda\log Z_2$, is negative.

As mentioned in the Introduction, Euclidean volume fluctuations might correspond to fluctuations of the cosmological horizon. In $dS_4$, the distance to the horizon is proportional to $V_{cl}^{1/4}$. This suggests that the fluctuations of ${\mathcal V}^{1/4}$ might tell us something about the fluctuations of the horizon distance. These fluctuations can be estimated as follows.
Since, with the actual values of $G$ and $\Lambda$, the probability distribution for the volume is very much concentrated around the classical value, for a generic function $f$ we have
\begin{equation}
    \var[f({\mathcal V})]\approx\var[f(V_{cl})+f'(V_{cl})({\mathcal V}-V_{cl})]=[f'(V_{cl})]^2\var({\mathcal V})\sim [f'(V_{cl})V_{cl}]^2 G\Lambda,
\end{equation}
where in the last step we used the leading-order result (\ref{varsaddle}) for the variance of the volume and that, in 4 dimensions, $V_{cl}\propto\Lambda^{-2}$. Therefore, the relative fluctuations of any power of ${\mathcal V}$ are $\sqrt{\var({\mathcal V}^\alpha)}/V_{cl}^\alpha\sim\sqrt{G\Lambda}\sim 10^{-61}$, a tiny number.

Let us make some comments on the case $d=2$, where instead of the volume it is more natural to speak of the area. The probability distribution (\ref{prob2d}) is a normalized version of the fixed-area partition function, commonly considered in the context of 2D quantum gravity (see for example \cite{seiberg_1990_notes,ginsparg_2024_lectures}). Being normalized makes it much easier to compute, because it only requires knowing the dependence of $Z$ on the cosmological constant, and $\Lambda$-independent factors are irrelevant. In fact, the value of the timelike Liouville partition function (the main piece of the 2D gravity partition function) is under some debate \cite{anninos_2022_quantum,giribet_2022_2d,collier_virasoro,collier_complex,giribet_2025_deforming}, but its dependence on the cosmological constant is robustly given by KPZ scaling. 
On the other hand, the probability distribution we obtained in the 2D case has a peculiar feature: it becomes a delta function centered at the origin, $p(A)=\delta(A)$, when the matter central charge takes its minimum possible value, $c_m=25$ (which we may view as the ``most quantum'' regime, because it is opposite to the classical limit, $c_m\to\infty$). This result does not seem to change when one takes into account the effects of higher-genus topologies. What does this mean? The theory with $c_m=25$ only allows for microscopic universes? It would be interesting to have a physical explanation of why this is happening.

We close with some comments regarding holography in de Sitter space. Having a dual description of de Sitter gravity would mean one is able to reproduce gravitational observables from a non gravitating theory. In de Sitter space the situation is much less clear than in AdS, for example it is not clear if the dual theory should even be a unitary quantum mechanical theory \cite{Anninos:2011ui}. Nevertheless, the moments defined in \eqref{variance} and the probability distribution $p(V)$ are well defined observables of the gravitational theory, and therefore they would be good objects to try to reproduce from any proposed dual theory. Being able to do that might shed light into the physical interpretation of this probability distribution. In particular we believe that being able to reconstruct this $p(V)$ from a dual model should be easiest in the two dimensional case, in which we have a non perturbative answer for the sphere contribution \eqref{probliouville}. Moreover, at least in the context of spacelike Liouville there exists a relationship between fixed area quantities and matrix models, in which the Liouville computations can be understood as counting random surfaces \cite{Zamolodchikov:1982vx,Anninos:2020ccj,Anninos:2021eit,Anninos:2020geh}.\footnote{We thank Dionysios Anninos for pointing this out to us.} It would be interesting to see if the $p(A)$ that we obtained can be understood in similar terms. We leave this analysis for future work.

\acknowledgments

The authors thank Gaston Giribet, Juan Maldacena, Nicolás Kovensky and the rest of participants of the XXVII Giambiagi School for useful discussions. We are especially grateful to Themistocles Zikopoulos for a careful reading of the manuscript. This work has
been supported by UBA and CONICET and through the grants PICT 2021-00644, PIP
112202101 00685CO and UBACYT 20020220400140BA. DB also acknowledges support
from CONICET through the grant PIBAA 2872021010 0958CO. BS is supported
by a CONICET PhD fellowship. Research at Perimeter Institute is supported in part by the Government of Canada through the Department of Innovation, Science and Economic Development and by the Province of Ontario through the Ministry of Colleges, Universities, Research Excellence and Security.

\bibliographystyle{JHEP}
\bibliography{refs}

\end{document}